\documentclass[twocolumn,showpacs,prb]{revtex4}
\usepackage{graphicx,amsmath}

\newcommand{\RM}[1]{\MakeUppercase{\romannumeral #1{}}}
\begin{document}
\title{Quantum dot admittance probed at microwave frequencies with an on-chip resonator}
\author{T.~Frey}
\email{freytob@phys.ethz.ch}
\affiliation{Department of Physics, ETH Zurich, CH-8093 Zurich, Switzerland.}
\author{P.~J.~Leek}
\affiliation{Department of Physics, ETH Zurich, CH-8093 Zurich, Switzerland.}
\author{M.~Beck}
\affiliation{Department of Physics, ETH Zurich, CH-8093 Zurich, Switzerland.}
\author{J.~Faist}
\affiliation{Department of Physics, ETH Zurich, CH-8093 Zurich, Switzerland.}
\author{A.~Wallraff}
\affiliation{Department of Physics, ETH Zurich, CH-8093 Zurich, Switzerland.}
\author{K.~Ensslin}
\affiliation{Department of Physics, ETH Zurich, CH-8093 Zurich, Switzerland.}
\author{T.~Ihn}
\affiliation{Department of Physics, ETH Zurich, CH-8093 Zurich, Switzerland.}
\author{M.~B\"{u}ttiker}
\affiliation{D\'{e}partement de Physique Th\'{e}orique, Universit\'{e} de Gen\`{e}ve, 24 quai Ernest Ansermet, CH-1211 Gen\`{e}ve, Switzerland.}
\pacs{}

\date{\today}
\begin{abstract}
We present microwave frequency measurements of the dynamic admittance of a quantum dot tunnel coupled to a two-dimensional electron gas. The measurements are made via a high-quality 6.75 GHz on-chip resonator capacitively coupled to the dot.
The resonator frequency is found to shift both down and up close to conductance resonance of the dot corresponding to a change of sign of the reactance of the system from capacitive to inductive. The observations are consistent with a scattering matrix model. The sign of the reactance depends on the detuning of the dot from conductance resonance and on the magnitude of the tunnel rate to the lead with respect to the resonator frequency. Inductive response is observed on a conductance resonance, when tunnel coupling and temperature are sufficiently small compared to the resonator frequency.
\end{abstract}
\maketitle
\section{Introduction}
The tunneling of a particle through a potential barrier is one of the most fundamental processes in quantum mechanics.\cite{Sakurai1994} Research on transport experiments in semiconductor quantum dots \cite{kouwenhoven1997,Reimann2002} relies on the tunneling effect as a requirement to measure a charge current through these devices. The transmission of the tunnel barriers themselves can be investigated in detail over orders of magnitude e.g.~by counting electrons entering and leaving the quantum dot with a nearby charge detector. \cite{Lu2003,Gustavsson2009} The properties of quantum dots coupled via a tunnel barrier to the leads can be studied not only with direct current (DC) bias, \cite{Gustavsson2008} but also at finite frequency, measuring the complex admittance, \cite{Gabelli2006,Delbecq2011,Ciccarelli2011} giving information about the dynamics of the electrons in the system.
Previous experiments at frequencies smaller than tunnel coupling and temperature have shown good correspondence to theory when the quantum dot circuit is modeled as a capacitor and a resistor connected in series. \cite{Gabelli2006} A theoretical treatment of this dynamic admittance using a scattering matrix approach \cite{Nazarov2009} has led to the proposal that for large ratios between excitation frequency and tunnel rate, inductive as well as capacitive behavior should be observable in these systems. \cite{Buttiker1996} Such effects are expected to be more pronounced at higher frequencies.\cite{Wang2007}

In this paper we present measurements of the finite frequency response of a single quantum dot for different tunnel rates to a two-dimensional electron gas (2DEG). In particular, we investigate in detail the dynamic admittance of a double quantum dot arising between one dot and its lead when the other dot is detuned such that its influence can be neglected. These measurements are in a different regime in comparison to those in Ref.~\onlinecite{Frey2012}, where only the dynamic coupling between the two dots and microwave photons was investigated and the coupling to the leads was irrelevant.
The capacitive coupling to an on-chip high-quality transmission line resonator is used for the measurement. Such resonators have been shown to be excellent devices for realizing strong coupling cavity quantum electrodynamics \cite{Wallraff2004} and quantum non-demolition readout \cite{Wallraff2005,Filipp2009,Reed2010} of superconducting qubits. We find that in the vicinity of a conductance resonance \cite{kouwenhoven1997} between the lead and the dot, the resonant frequency of the microwave resonator shifts both up and down in frequency, depending on the dot-lead tunnel rate. Our results allow us to address a long-standing theoretical concept on inductive and capacitive effects in the admittance of mesoscopic samples \cite{Buttiker1996} in a parameter regime that has not been explored before to the best of our knowledge. The results and their interpretation allow us to give an intuitive picture for the appearance of the inductive effect. 

\section{Sample and setup}
\begin{figure}[htp]
	\centering
		\includegraphics[width=1.00\columnwidth]{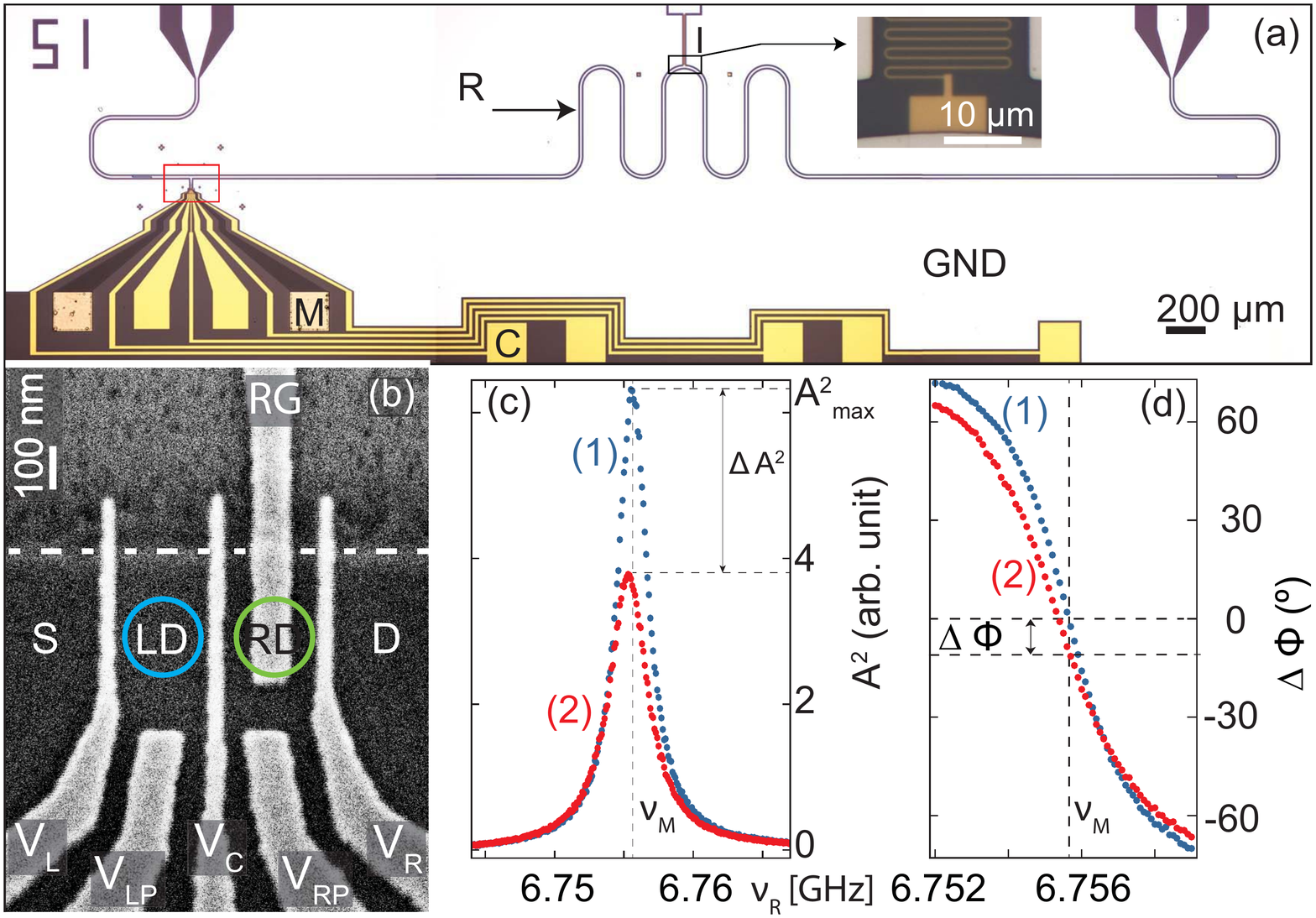}
	\caption{(Color online) (a) Optical image of the microwave resonator-double quantum dot sample. Ohmic contacts (M), top gates (C), resonator (R), ground plane (GND) and on-chip inductor (inset, I). (b) Scanning electron micrograph of the double quantum dot structure (LD, RD) defined by the gates $V_{\rm{L}}$, $V_{\rm{LP}}$, $V_{\rm{C}}$, $V_{\rm{RP}}$, $V_{\rm{R}}$. RG marks the gate connected to the resonator. The mesa edge is highlighted by a dashed line. The double quantum dot is connected to the 2DEG source (S) and drain (D) contacts. (c,d) Dependence of the transmission amplitude (c) and phase (d) of the resonator on the frequency $\nu_{\rm{R}}$ of the applied microwave signal in Coulomb blockade (1) and on a conductance resonance (2).}
	\label{Fig1}
\end{figure}

The sample used in our experiments is shown in Fig.~\ref{Fig1}(a) (see also Ref.~\onlinecite{Frey2012}) and consists of a superconducting microwave resonator \cite{Goeppl2008} coupled to a double quantum dot. The resonator is made of aluminum and realized as a coplanar waveguide.\cite{Simons2002} It is capacitively coupled via an extension [resonator gate (RG)] reaching from the center conductor of the resonator to the right quantum dot (RD) [Fig.~\ref{Fig1}(b)]. This ensures a much larger capacitive coupling of the resonator to the right dot than to the left dot or the leads. 

The double quantum dot is defined on a mesa etched into the 2DEG chip. The 2DEG is realized in a $\rm{Al_{x}Ga_{1-x}As}$ heterostructure and forms $35~\rm{nm}$ below the surface. The edge of the mesa [dashed line in Fig.~\ref{Fig1}(b)] is the upper boundary of the double quantum dot. The confinement of charges in the dot is realized by applying appropriate negative voltages to the gates. In this way, the left-, right- and center gate voltages ($V_{\rm{L}}$, $V_{\rm{R}}$, $V_{\rm{C}}$) predominantly control the tunnel barriers to the two 2DEG leads [source (S), drain (D)] and between the dots, respectively [Fig.~\ref{Fig1} (b)]. The two plunger gates ($V_{\rm{LP}}$, $V_{\rm{RP}}$) are used to change the number of electrons in the dots. Due to finite capacitive cross coupling, the tunnel barrier gates also influence the number of electrons in the dots. 

The sample is mounted in a dilution refrigerator to perform measurements at a base temperature of around $10~\rm{mK}$. We extract an electron temperature $T_{\rm{e}}\leq 135~\rm{mK}$ from a fit to a thermally broadened conductance resonance. \cite{kouwenhoven1997} With a setup similar to the one described in detail in Ref.~\onlinecite{Frey2012}, we perform current measurements in the sub-kHz regime and we measure the transmission of a microwave signal through the resonator. A heterodyne detection scheme is used, following the ideas developed in circuit QED experiments,\cite{Wallraff2004} to determine the two quadratures of the transmitted microwave field and therefore its amplitude and phase.
\section{Measurements}
A Lorentzian resonance [Fig.~\ref{Fig1} (c) and (d)] is obtained when recording the transmission through the microwave resonator as a function of the applied microwave frequency $\nu_{\rm{R}}$. We extract a resonant frequency $\nu_{\rm{res}} \approx 6.755~\rm{GHz}$ and a loaded quality factor \cite{Goeppl2008} $Q_{\rm{L}} \approx 2610$ for the fundamental mode of the resonator with all quantum dot gates grounded. 

\begin{figure*}[htp]
	\centering
		\includegraphics[width=1.00\textwidth]{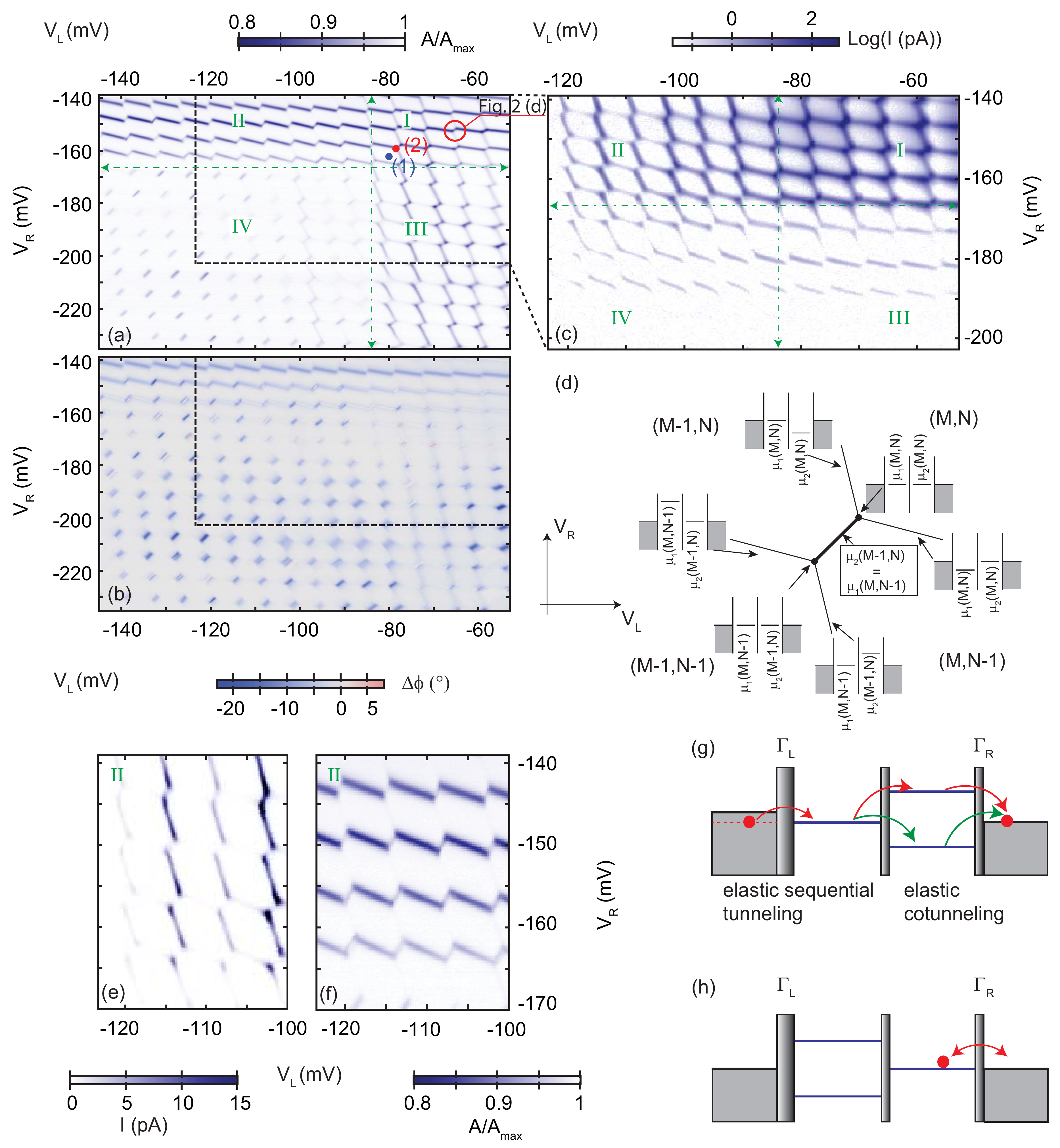}
	\caption{(Color online) (a) Relative resonator transmission amplitude $A/A_{\rm{max}}$ at fixed measurement frequency $\nu_{\rm{M}}$ as a function of $V_{\rm{L}}$ and $V_{\rm{R}}$. Points labeled (1) and (2) mark gate voltage positions where the full resonance spectra shown in Fig.~\ref{Fig1} (c,d) were measured. Green roman numerals label four different measurement regions separated by dash-dotted green lines, details in the text. (b) Phase change of the microwave resonator for the same gate voltage ranges as in (a). (c) Direct current measurement through the double quantum dot for gate voltage settings $V_{\rm{L}}$ and $V_{\rm{R}}$ within the dashed region highlighted in (a). (d) Schematic of the charging diagram of a double quantum dot for (N,M) electrons close to the two triple points as a function of $V_{\rm{L}}$ and $V_{\rm{R}}$. (e) Zoom of the direct current measurement in region \RM{2}. (f) Magnified view of the relative transmission amplitude measurement in the same gate voltage range as in (e). (g,h) Schematic of the double quantum dot for the direct current measurement (g) and for the microwave measurement (h), respectively.}
	\label{Fig2}
\end{figure*}

We first investigate the dependence of the quality factor and resonant frequency of the resonator on the charge configuration when forming the double quantum dot. The transmission amplitude and phase are displayed in Fig.~\ref{Fig1}(c,d) for two different gate voltage settings. For these and all the other microwave measurements shown in this paper the source and drain contacts are grounded. For the first case marked (1) in Fig.~\ref{Fig1}(c,d) the quantum dot is tuned into Coulomb blockade, \cite{Nazarov2009} where the number of electrons is fixed in both dots. For the second case (2), the double quantum dot is tuned to a triple point. Here the chemical potential in the leads and a state in each dot are aligned such that it is energetically possible for an electron to propagate elastically from one lead through the quantum dot into the other lead. The gate voltage settings of the two measurements (1) and (2) are labeled in Fig.~\ref{Fig2} (a).  A reduction in amplitude and a change in phase is observed between the two cases while the line shapes remain Lorentzian [Fig.~\ref{Fig1}(c,d)]. This fact can be exploited to save measurement time; instead of measuring the full resonance spectrum, a microwave tone at a fixed frequency $\nu_{\rm{M}}=\nu_{\rm{res}}$ is applied to the resonator and the difference in amplitude  $\Delta A$ and in phase $\Delta \phi$ is recorded.

In Fig.~\ref{Fig2} the change in amplitude (a) and phase (b) are shown for several different electron configurations in the two dots. No major undesired background charge rearrangement is visible over the whole investigated gate voltage region. In Fig.~\ref{Fig2} (d) the sketch of a double dot charging diagram \cite{vanderWiel2003} is shown along with the corresponding quantum dot level structure in the vicinity of two triple points, as indicated in Fig.~\ref{Fig2} (a). We can identify different characteristic regions by taking a closer look at Fig.~\ref{Fig2} (a) and using this schematic. In region \RM{1} [Fig.~\ref{Fig2} (a)], charge transfer between the two dots and between each dot and its adjacent lead is observed. In region \RM{2} (\RM{3}) only charge transfer to the right (left) lead is discernible together with interdot charge transfer.
The resonances to the right lead are more pronounced than those to the left lead. This is expected because the resonator is more strongly capacitively coupled to the right quantum dot than to the left, leading to a higher sensitivity to the tunnel coupling between the right dot and its lead. In region \RM{4}, no charge transfer between the dots and the leads are visible. Microwave measurements in this region are discussed in detail in Ref.~\onlinecite{Frey2012}.

In this paper, we focus on region \RM{2} in which we specifically study the interaction between the right dot and the right lead. Fig. \ref{Fig2} (b) shows that not only a change in amplitude is detected when a state in the right dot is resonant with the lead but also a change in phase. Since the measurement frequency is set to $\nu_{\rm{res}}$, such a change in phase corresponds to a change in resonance frequency. This frequency shift can be caused by reactive as well as dissipative changes in the impedance of the right dot tunnel-coupled to its lead, discussed in more detail below.
 
We compare the microwave frequency response with standard DC transport measurements \cite{kouwenhoven1997} as shown in Fig.~\ref{Fig2}~(c) to understand the origin of the finite frequency response of the quantum dot-lead circuit. Voltages $V_{\rm{L}}$ and $V_{\rm{R}}$ are swept as in Fig.~\ref{Fig2} (a,b), but now a small source-drain bias $V_{\rm{SD}}\approx 50~{\rm{\mu V}}$ is applied. In Fig.~\ref{Fig2} (c) clear hexagon patterns are observable. The size of the hexagons and their position in gate voltage are the same as in the microwave measurements. For less negative gate voltages, the coupling to the leads is strong enough to lead to a finite current within the hexagons indicating that the dots are not deep in the Coulomb blockade regime. With decreasing gate voltages $V_{\rm{L}}$ and $V_{\rm{R}}$, and therefore decreasing tunnel rates, the resonances to the leads become smaller and finally disappear. 

Although the microwave and transport measurements show equivalent charging diagrams, differences are observable, such as good or poor visibility of the interdot charging line and of the resonances to the different leads. This indicates that the physical origin of the signal is different for the two measurement techniques. We discuss this aspect in more detail below.

In the DC measurement, the interdot charging lines at which $\mu_{\rm{1}}(M,N-1)=\mu_{\rm{2}}(M-1,N)$ [Fig.~\ref{Fig2} (d)] are not observed because here transport is suppressed due to Coulomb blockade.\cite{vanderWiel2003} 
 
Concerning the dot-lead resonances, sketches visualizing a possible explanation for the difference in the signals are shown in Fig.~\ref{Fig2} (g,h). The transport data in region \RM{3} [Fig.~\ref{Fig2}(c)] shows mainly cotunneling lines \cite{vanderWiel2003} with the right dot being in resonance with the lead with decreasing $V_{\rm{R}}$. In the corresponding microwave measurement however, the resonances to the left lead are more pronounced. Equivalent features can also be found in region \RM{2} with decreasing $V_{\rm{L}}$, as shown in Fig.~\ref{Fig2} (e,f). This time the resonance to the left lead is seen in the transport data and the resonance to the right lead in the microwave measurements. This indicates that the microwave signal strength for a dot state being resonant with a lead depends strongly on the tunnel rate to this lead [Fig.~\ref{Fig2} (h)].

For the transport measurements however, an electron must pass through both quantum dots for a current to be measured [Fig.~\ref{Fig2} (g)]. For sufficiently negative left side gate voltages ($V_{\rm{L}} \lesssim -100~\rm{mV}$), a cotunneling current can only be observed along two of the boundaries of the hexagon [Fig.~\ref{Fig2} (e)]. Along these lines the tunnel rate to the left dot is small, but since a dot level is resonant with the lead, first order tunneling is possible and the electron can enter the left dot. In order to now pass through the right dot, a cotunneling process needs to occur which requires a strong coupling between the right dot and lead.\cite{IhnBuch}

We conclude that the dominant mechanism for a change in the microwave transmission signal is the tunneling process of an electron back and forth between the lead and the dot, if only one of the dots is resonant with a lead and if the two dots are tuned more than the energy of a microwave photon apart, to exclude interdot photon assisted tunneling. \cite{Oosterkamp1998}

\begin{figure}
	\centering
		\includegraphics[width=1.00\columnwidth]{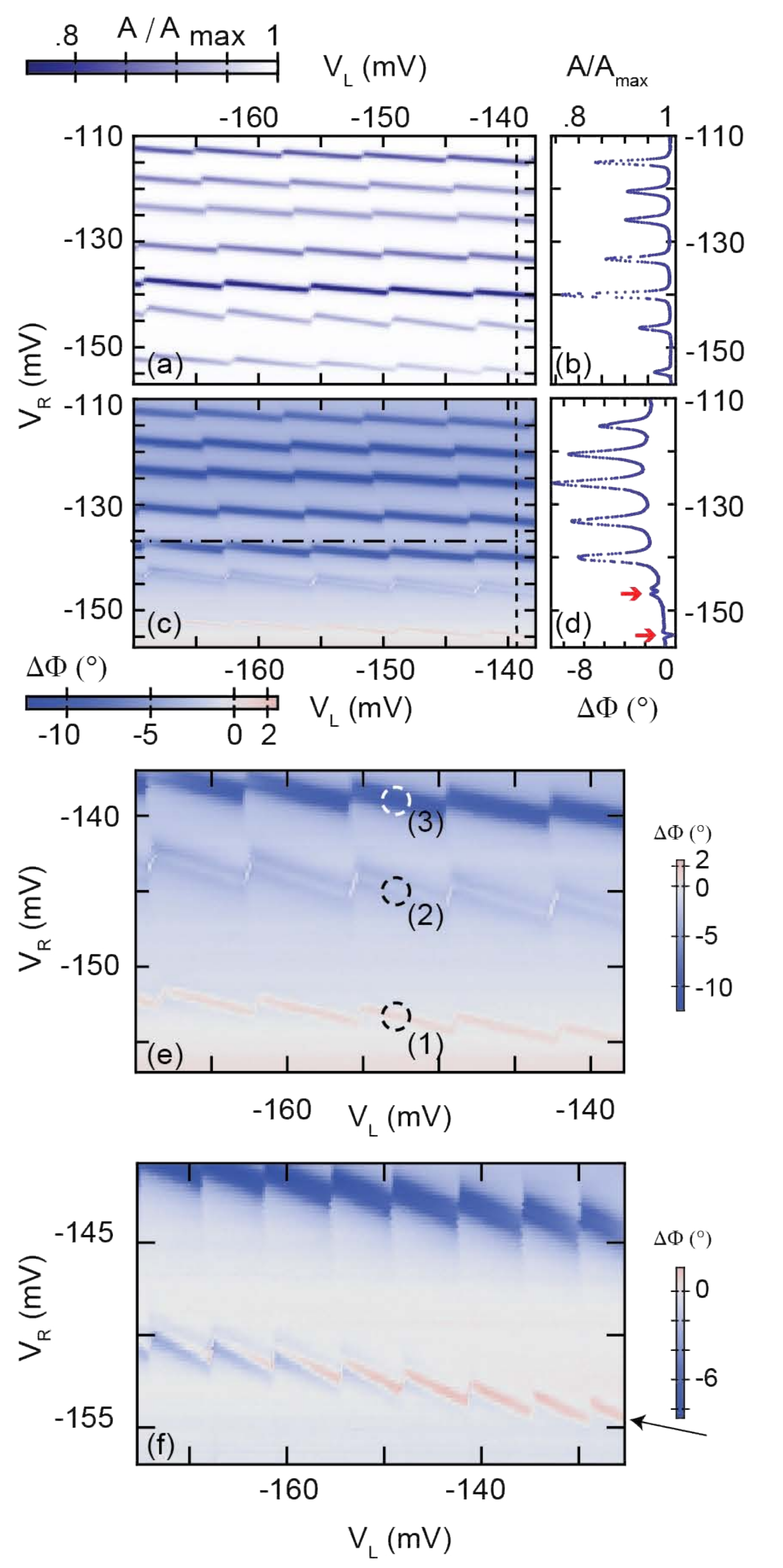}
	\caption{(Color online) (a) Relative resonator transmission amplitude $A/A_{\rm{max}}$ at fixed measurement frequency as a function of $V_{\rm{L}}$ and $V_{\rm{R}}$. (b) Relative amplitude change versus $V_{\rm{R}}$ along the dashed line shown in (a). (c) Phase change $\Delta \phi$ for the same gate voltage range as shown in (a). (d) Change of transmission phase for a fixed measurement frequency $\nu_{\rm{M}}$ along $V_{\rm{R}}$ as indicated in (c). (e) Detail of the transmission phase in a selected region of $V_{\rm{R}}$ as indicated by a dash-dotted line in (c). Dashed circles refer to gate voltage settings investigated in more detail in Fig.~\ref{Fig4}. (f) Change of transmission phase $\Delta \phi$ at fixed $\nu_{\rm{M}}$ as a function of $V_{\rm{R}}$ and $V_{\rm{L}}$ for a dataset acquired in a similar regime, but with slightly different gate voltages necessary after a charge rearrangement in the sample. }
	\label{Fig3}
\end{figure}

We have further investigated the response of the resonator close to dot-lead resonances (see Fig.~\ref{Fig3}). We now operate $V_{\rm{L}}$ and $V_{\rm{C}}$ at more negative gate voltages than in Fig.~\ref{Fig2}, in order to focus on tunneling events occurring between the right lead and the right quantum dot. This results in smaller tunnel rates to the left lead and between the dots. Whenever one of the states in the right dot is resonant with the lead, a drop in transmission amplitude is visible [\ref{Fig3} (a,b)]. We see more interesting features however in the transmission phase [\ref{Fig3} (c,d)], in which a characteristic change in the response occurs with decreasing $V_{\rm{R}}$ (i.e.~decreasing tunnel rate to the lead). At first the change in phase $\Delta \phi$ is negative, indicating that the resonance frequency shifts to lower values. But at $V_{\rm{R}} \approx -145~\rm{mV}$, we observe a double peak in the phase response and for even smaller tunnel rates, the phase response becomes positive (highlighted in Fig.~\ref{Fig3} (d) by red arrows).
This characteristic change in the phase response is seen for a range of different electron numbers in the left dot [Fig.~\ref{Fig3} (e)]. A further dataset indicating that this change in the phase signal depends on the tunnel rate between the right dot and the lead is shown in Fig.~\ref{Fig3} (f). This is a dataset acquired in a similar regime, but the gate voltage parameters have slightly changed compared to Fig.~\ref{Fig3} (e) due to a charge rearrangement in the system. Here the positive phase shift of the resonator evolves into a negative phase shift as the same dot-lead resonance is followed to more positive  $V_{\rm{R}}$, corresponding to an increasing tunnel rate (indicated by an arrow).

We have found that the spacing of the double peak in the phase signal does not depend on the microwave power over several orders of magnitude. We have also confirmed experimentally that the origin of the double peak in the phase signal is not caused by a current rectification process, as follows. We have measured the phase response with a low-frequency sine signal added to the lead or the resonator gate. As a result, the double peaked phase signal was smeared out and for increasing amplitude of the sine signal a double peak in the transmission amplitude of the resonator was observed (data not shown).

We now discuss three different characteristic measurements of the phase response in more detail for the gate voltages highlighted by dashed circles in Fig.~\ref{Fig3} (e). For these datasets there is a nontrivial dependence of the frequency shift on the gate voltage $V_{\rm{R}}$. We have measured the full microwave transmission spectrum when tuning the right quantum dot through resonance with the lead for these three voltage settings. We find that for all measurements, the transmission amplitude can be fitted with a Lorentzian and we obtain the resonance frequency, which for small shifts $\Delta \nu\ll \kappa/2\pi$ is directly proportional to the change in the transmission phase. In Fig. \ref{Fig4} (d) we plot the resonant frequency shift $\Delta \nu_{\rm{0}}$ versus the change in right side gate voltage, where the resonances have been horizontally offset for better comparison. A small linear background phase [visible in Fig.~\ref{Fig3} (d)] has been subtracted from the data in Fig.~\ref{Fig4} (d), and is likely to originate from a slight change in direct capacitance between the resonator and the right lead.
\section{Interpretation}

\begin{figure*}[htbp]
	\centering
		\includegraphics[width=1.00\textwidth]{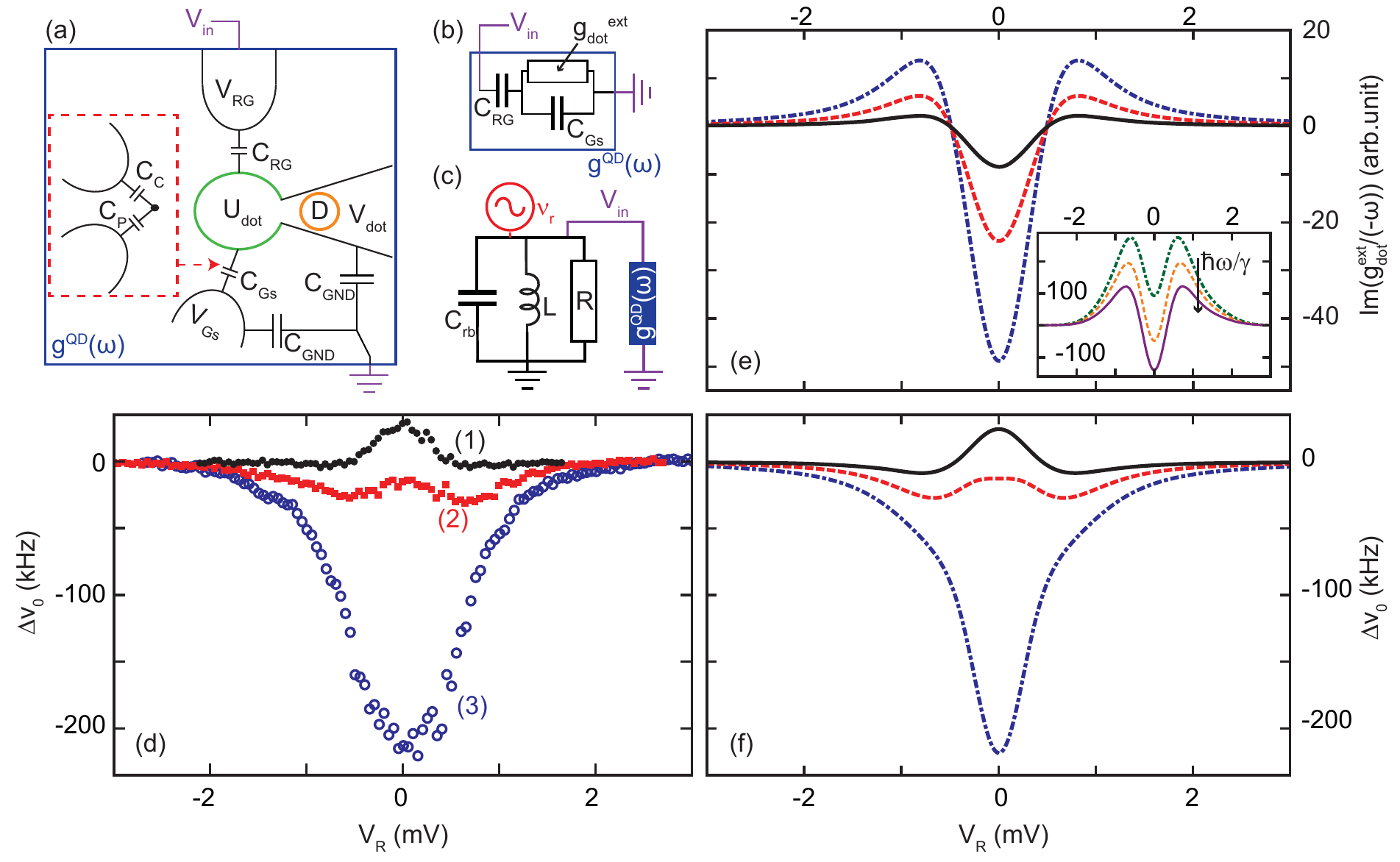}
	\caption{(Color online) (a) Schematic of the right quantum dot tunnel-coupled to the drain lead (D). The dot is capacitively coupled to the resonator gate (RG) and to a capacitor ($C_{\rm{Gs}}$) formed by the capacitive coupling of the center gate ($C_{\rm{C}}$) and the right plunger gate ($C_{\rm{P}}$). The displayed model circuit is used to calculate the dynamic admittance $g^{\rm{QD}}(\omega)$ of the dot coupled to the lead at finite frequencies. The $V_{\rm{x}}$ refer to the voltages applied by external voltage sources with $\rm{x}=\rm{dot}$, $\rm{Gs}$, $\rm{RG}$. $U_{\rm{dot}}$ describes the locally induced potential in the quantum dot. (b) Circuit representation of the schematic in (a), see Eq.~(\ref{Buttikerscattering}) in the text. (c) Lumped element model of the microwave resonator with the dynamic admittance of the quantum dot connected in parallel. (d) Measured change of resonator frequency $\Delta \nu_{\rm{0}}$ for the three resonant dot lead states (1-3) indicated by dashed circles in Fig.~\ref{Fig3} (e). (e) Calculated imaginary part of complex external admittance $g^{\rm{ext}}_{\rm{dot}}$ of the quantum dot [Eq.~(\ref{cadmit})] for tunnel couplings $\gamma_{\rm{1}}/h=20~\rm{MHz}$, $\gamma_{\rm{2}}/h=58~\rm{MHz}$, $\gamma_{\rm{3}}/h=125~\rm{MHz}$. Inset: Imaginary part of the external admittance for tunnel couplings $\gamma_{\rm{4}}/h=2~\rm{GHz}$, $\gamma_{\rm{5}}/h=1.5~\rm{GHz}$, $\gamma_{\rm{6}}/h=1~\rm{GHz}$ from top to bottom. (f) Results of the model calculations [Eq.~(\ref{nurescal})], for three different tunnel couplings from top to bottom as in (e).}
	\label{Fig4}
\end{figure*}

In order to understand the observed frequency shifts of the resonator at different tunnel rates, we model the system using the scattering matrix approach described in Ref.~\onlinecite{Buttiker1996}.  
We neglect the presence of the left quantum dot in our model. In addition, as the capacitance between the right plunger gate and the center gate and their capacitance to ground are much larger than the individual capacitances to the dot, we assume that they are always on the same microwave potential. This allows us to model the capacitive coupling of these two gates to the right dot by one effective gate with a capacitance $C_{\rm{Gs}}$ [Fig.~\ref{Fig4} (a)]. In the model circuit, depicted in Fig.~\ref{Fig4} (a), the resonator gate is coupled via a capacitor $C_{\rm{RG}}$ to the quantum dot and the drain contact (D) is tunnel coupled to the dot. The voltages $V_{\rm{x}}$ ($\rm{x}=\rm{dot}$, $\rm{Gs}$, $\rm{RG}$) are applied with external sources, and the voltage $U_{\rm{dot}}$ stands for the internal electrostatic potential of the quantum dot. This quantity arises due to the Coulomb interaction between charged particles in the dot, the lead and the gates and results in a local potential energy of electrons on the quantum dot.\cite{Buttiker1996} The scattering matrix can be further simplified by approximating the density of states in the metal leads as being infinite, since the density of states in a metal is much larger than the density of states in the quantum dot. \cite{Wang2007} We calculate the dynamic admittance $g^{\rm{QD}}(\omega)$ of the quantum dot circuit in this approximation and find for the current flowing into the resonator gate:\cite{fouriertrafo}
\begin{multline}
I_{\rm{RG}}=\left(\frac{1}{-i \omega C_{\rm{RG}}}+\frac{1}{g^{\rm{ext}}_{\rm{dot}}(\omega)-i\omega C_{\rm{Gs}}}\right)^{-1} \cdot V_{\rm{RG}}\\
 = g^{\rm{QD}}(\omega) \cdot V_{\rm{RG}}.
\label{Buttikerscattering}
\end{multline}
The lumped element representation of $g^{\rm{QD}}(\omega)$ [Eq.~(\ref{Buttikerscattering})] is shown in Fig. \ref{Fig4} (b). It consists of the parallel circuit of the capacitor $C_{\rm{Gs}}$ and the complex admittance of $g^{\rm{ext}}_{\rm{dot}}(\omega)$ in series with the coupling capacitor $C_{\rm{RG}}$. 
In Eq.~(\ref{Buttikerscattering}),
\begin{multline}
g^{\rm{ext}}_{\rm{dot}}(\omega,\mu,kT,\gamma)=\\
\frac{e^2}{h}\int d\epsilon [1-s^{\dagger}(\epsilon ,\gamma) s(\epsilon +\hbar \omega ,\gamma )]
\left(\frac{f(\epsilon)-f(\epsilon +\hbar \omega)}{\hbar \omega }\right)
\label{cadmit}
\end{multline}
describes the external response of the quantum dot coupled to a single lead with an excitation with frequency $\omega/2\pi$. The two functions in $g^{\rm{ext}}_{\rm{dot}}$ are the scattering matrix $s(\epsilon,\gamma)=(1+i \epsilon /\gamma)/(1-i \epsilon /\gamma )$, describing resonant reflection of Breit-Wigner type with $\gamma$ being the tunnel coupling strength, and the Fermi distribution $f(\epsilon,\mu,kT)=1/\{\exp[(\epsilon -\mu )/kT]+1\}$. Here, $\mu$ describes the energetic detuning of the quantum dot level and the Fermi energy in the lead. It translates into a side gate voltage via $\mu=\alpha_{\rm R} V_{\rm R}$, where $\alpha_{\rm R}$ is the lever arm of the right side gate.
The external response $g^{\rm{ext}}_{\rm{dot}}(\omega,\mu,kT,\gamma)$ can be rewritten in terms of energy ratios $\hbar\omega/\gamma$, $\hbar\omega/kT$ and $\mu/kT$. Its functional form is therefore completely determined by these three values.
We assume that the source contact and other gates are at microwave ground as $C_{\rm{GND}}$ is much larger than any other capacitances and the dot is only modulated by the gate connected to the resonator.

We also map the resonator to a lumped element model, \cite{Goeppl2008} to obtain the model circuit shown in Fig.~\ref{Fig4} (c). To investigate the frequency shift $\Delta \nu$ of the resonator 
\begin{equation}
\Delta \nu=\frac{1}{4 \cdot 50~\Omega(C_{\rm{rb}}+Y_{\rm{QD}})}-\nu_{\rm{offset}}
\label{nurescal}
\end{equation}
we use $Y_{\rm{QD}}=\rm{Im}[g^{\rm{QD}}(\omega)]/(-\omega)$ and $\nu_{\rm{offset}}$ is a constant offset subtracted to obtain $\Delta \nu$.

Since the results of Eq.~(\ref{Buttikerscattering}),~(\ref{cadmit}),~(\ref{nurescal}) depend on a number of parameters ($C_{\rm{RG}}$, $C_{\rm{Gs}}$, $\hbar \omega$, $\gamma$, $kT$, $C_{\rm{rb}}$, $\alpha_{\rm R}$), we fix as many parameters as possible by using values which are either directly extracted from experimental data or are reasonable, given the boundary conditions. 
We use $C_{\rm{RG}}=56~\rm{aF}$ and $C_{\rm{Gs}}=30~\rm{aF}$ obtained from gate characterization measurements, and $C_{\rm{rb}}\approx 0.74~\rm{pF}$, calculated from the resonator impedance and frequency. For the electron temperature a value of $135~\rm{mK}$ is taken and $\omega=2 \pi \nu_{\rm{res}}$. A constant lever arm $\alpha_{\rm R}=0.05$ is used to convert the energy scale to a gate voltage, which is within less than a factor of two equal to the value obtained from finite bias measurements in another regime. The tunnel coupling $\gamma$ is the only adjustable parameter. It is chosen to match the different data sets best and is in agreement with an upper bound obtained from the DC pinch off curve of the right side gate with the center gate and the left side gate open. 

To compare the results of Eq.~(\ref{nurescal}) with the data [Fig.~\ref{Fig4} (d)], we first show the complex admittance $g^{\rm{ext}}_{\rm{dot}}(\omega)$ for different ratios between the energy of a microwave photon and the tunnel coupling $\hbar \omega /\gamma$. In the inset of Fig.~\ref{Fig4} (e),  the reactance $\rm{Im}[g^{\rm{ext}}_{\rm{dot}}(\omega)]/(-\omega)$ is plotted for increasing values of $\hbar \omega /\gamma$. For the topmost curve, the response is capacitive for all gate voltages $V_{\rm{R}}$. As the tunnel coupling is decreased, inductive behavior begins  to develop close to resonance ($V_{\rm{R}}\approx 0$). The inductive response is strong in the regime where $kT \ll \hbar \omega$ and $\gamma \ll \hbar \omega$. It starts to appear at $\gamma \approx \hbar \omega$ and $\gamma \gtrsim kT$.

The main part of Fig.~\ref{Fig4} (e) shows the external response of the dot-lead system for parameters as used for the calculation of the resonator frequency shift in Fig.~\ref{Fig4} (f). In all three cases an inductive behavior is observed close to resonance which turns into a capacitive response with detuning from the resonance.

The results of the evaluation of $\Delta \nu$ using Eq.~(\ref{nurescal}) are shown in Fig.~\ref{Fig4} (f). We compare the three different data sets in Fig.~\ref{Fig4} (d) with these calculations and observe that the characteristic features of our measurements are well reproduced within this model. We find a dynamic admittance $g^{\rm{QD}}(\omega)$ with only a capacitive component for the most transparent tunnel barrier of the three measurement sets presented in Fig.~\ref{Fig4}. The model also reproduces the crossover from a capacitive to an inductive behavior as the tunnel rate is reduced. For $g^{\rm{QD}}(\omega)$ the transition between inductive and capacitive behavior is shifted down from the GHz to the few tens of MHz-range due to the presence of the additional capacitances $C_{\rm Gs}$ and $C_{\rm RG}$. 

In the following we try to give an intuitive explanation for the appearance of an inductive and a capacitive response in the vicinity of a conductance resonance.
The decisive parameter for the external response to be inductive or capacitive close to resonance with the lead, is the ratio between the photon energy $\hbar \omega$ and the tunnel coupling $\gamma$.
As $\gamma$ decreases at fixed frequency $\omega$, the dwell time in the dot of a resonantly tunneling electron becomes longer. If this dwell time exceeds the period of the driving field, the electron can no longer follow the drive, the resulting current lags behind the voltage, and the response tends to be more and more inductive. This behavior may be interpreted as the quantum tunneling analogue of the kinetic inductance of an electron gas which is due to the inertia of the electrons causing the same lag between current and voltage. However, in our resonant structure this inductive effect competes with the well-known quantum capacitance contribution to the response caused by the tunneling broadened spectral density of the resonant state. The capacitive response prevails with increasing detuning from resonance, where tunneling of electrons into the dot has an increasing character of virtual transitions (elastic cotunneling) happening on very short time scales given by the detuning energy.

Comparing Figs.~\ref{Fig4} (d) and (f) we find the measured curve (3) to be slightly broader in gate voltage than the corresponding calculated trace. Also the capacitive reactance in case (1) is slightly less pronounced in the measurement than in the calculated curve.
Such differences between the scattering matrix model and the measurements could have a number of origins. Firstly the model does not consider any sources of decoherence and noise. Secondly the assumption of a Breit-Wigner phase dependence \cite{IhnBuch,Schuster1997}, which is based on a single resonance, may be an oversimplification.

Detailed analysis of the resonator linewidth gives access to dissipative effects described by the real part of the dynamic admittance. For dataset (2) we obtain good agreement between model and data (not shown) using the same parameters which describe the resonator frequency shift in Fig.~\ref{Fig4}. The model underestimates the dissipation by less than a factor 2 in the other two cases. However, a comparison of the measured phase and amplitude changes in Figs.~\ref{Fig3} (b) and (d) shows already that the dissipative effects may be harder to describe than the reactive effects. While the latter show a systematic and steady evolution from the case of purely capacitive response to the case of inductive response with decreasing gate voltage, the dissipative effects vary strongly in the same gate voltage range. While we do not have a clear explanation for this effect at present, we point out that charge relaxation, which is still under investigation,\cite{Buttiker2006,Mora2010,Filippone2012}  is believed to play an essential role for high frequency resistance measurements.  

\section{Conclusion}
In conclusion we have investigated the tunneling process between a quantized state and the continuum for different tunnel rates. Carrying out these measurements using a high frequency resonator enabled us to observe a crossover from a capacitive to an inductive response. The observations are well described within the scattering matrix model of Ref.~\onlinecite{Buttiker1996}. Our interpretation of the phase shift data is of particular relevance for similar experiments coupling a photon resonator to a double quantum dot \cite{Toida2012,petersson2012}. In particular, Ref. \onlinecite{Toida2012} reports similar phase shifts without discussion or interpretation. Our measurement scheme promises the feasibility of future experiments in which the sensitivity of the resonator is further exploited, e.g., in measuring the shot noise properties \cite{Reznikov1995,Kumar1996} of a tunnel barrier in more detail. 
\section*{Acknowledgments} 
We thank P.~Studerus, G.~Blatter, C.~R\"ossler and C.~Lang for technical support and discussions.
This work was financially supported by EU IP SOLID, by the Swiss National Science Foundation through the National Center of Competence in Research `Quantum Science and Technology', and by ETH Zurich.
\vspace*{-5mm}

\bibliographystyle{apsrev}
\end{document}